\input{aipcheck}

\documentclass[
    ,final            
    ,draft            
    ,numberedheadings 
  ]
  {aipproc}

\usepackage{amsmath}

\renewcommand{\vec}[1]{\mathrm{\bf #1}}

\layoutstyle{6x9}

\begin{document}

\title{\textbf{Cosmological consequences of Modified Gravity (MOG)}}

\author{Viktor T. Toth}{address={Ottawa, ON  K1N 9H5, Canada}}

\begin{abstract}
As an alternative to the $\Lambda$CDM concordance model, Scalar-Tensor-Vector Modified Gravity (MOG) theory reproduces key cosmological observations without postulating the presence of an exotic dark matter component. MOG is a field theory based on an action principle, with a variable gravitational constant and a repulsive vector field with variable range. MOG yields a phenomenological acceleration law that includes strong tensorial gravity partially canceled by a repulsive massive vector force. This acceleration law can be used to model the CMB acoustic spectrum and the matter power spectrum yielding good agreement with observation. A key prediction of MOG is the presence of strong baryonic oscillations, which will be detectable by future surveys. MOG is also consistent with Type Ia supernova data. We also describe on-going research of the coupling between MOG and continuous matter, consistent with the weak equivalence principle and solar system observations.
\end{abstract}

\maketitle


\section{Introduction}

Why is there a need for a modified gravity theory? There is a perfectly serviceable model of cosmology, the so-called $\Lambda$CDM ``concordance'' model, that is not only in good agreement with a large body of observational evidence, it also yielded some impressive predictions. Nonetheless, we feel motivated to seek alternatives, in part for the following reasons:
\begin{itemize}
\item The $\Lambda$CDM model requires 96\% of the universe to consist of black ``stuff'': cold dark matter and dark energy, both of which may never be detectable except through their gravitational influence;
\item The cold dark matter paradigm runs into difficulties even closer to home, notably its inability to explain convincingly why the rotation curves of spiral galaxies so closely follow their luminosity profiles.
\end{itemize}

The modified gravity theory we discuss here, Scalar-Tensor-Vector Gravity \cite{Moffat2006a,Moffat2007e} (STVG), has also been referred to by the acronym MOG more recently. MOG is a particularly interesting candidate for gravity modification in part because:
\begin{itemize}
\item In the solar system or the laboratory, MOG predicts Newtonian (or Einsteinian) physics;
\item The MOG acceleration law is consistent with star clusters \cite{Moffat2007a}, galaxies \cite{Brownstein2006a}, and galaxy clusters \cite{Brownstein2006b,Brownstein2007}.
\end{itemize}
In the rest of this presentation, we show that MOG also appears to be consistent with cosmological data \cite{Moffat2007c,Moffat2008a}. If these results hold, MOG may prove to be a more economical theory (in the sense of Occam's razor) than $\Lambda$CDM.

We begin with introducing MOG as a Lagrangian field theory in Section~\ref{sec:MOGfield}, also discussing the subject of coupling between the MOG fields and matter. Next, we briefly introduce the phenomenology of MOG, concentrating mostly on the spherically symmetric, static vacuum solution in Section~\ref{sec:MOGpheno}. We then move on to cosmology: we discuss the MOG prediction of the acoustic spectrum of the Cosmic Microwave Background (CMB) in Section~\ref{sec:MOGCMB} and the galaxy-galaxy matter power spectrum in Section~\ref{sec:MOGmatpow}. Finally, we move on to the topic that is the most active area of our current research, the study of MOG in the presence of continuous matter such as a perfect fluid, in Section~\ref{sec:MOGfluid}. We conclude with a brief discussion of the most significant challenges and outlook in Section~\ref{sec:end}.

\section{MOG as a field theory}
\label{sec:MOGfield}

MOG is a theory of gravity that augments Einstein's gravitational theory with a variable gravitational constant and a massive vector field with variable mass and coupling strength, producing a repulsive force. The theory's building blocks are:
\begin{itemize}
\item The tensor field $g_{\mu\nu}$ of metric gravity;
\item A scalar field $G$ representing a variable gravitational constant;
\item A massive vector field $\phi_\mu$ responsible for a repulsive force;
\item Another scalar field $\mu$ representing the variable mass of the vector field;
\item A further scalar field $\omega$ representing the variable coupling strength of the vector field.\footnote{Although $\omega$ is included for generality, in the solutions that we studied it turns out to be a constant.}
 \end{itemize}

MOG is a theory based on a Lagrangian action principle. The MOG Lagrangian has three parts: the Einstein-Hilbert Lagrangian of tensor gravity, the Lagrangian of the massive vector field, and the Lagrangian of the three scalar fields, complete with self-interaction potentials:
\begin{align}
{\cal L}=&-\frac{1}{16\pi G}\left(R+2\Lambda\right)\sqrt{-g}\\
&-\frac{1}{4\pi}\omega\left[\frac{1}{4}B^{\mu\nu}B_{\mu\nu}-\frac{1}{2}\mu^2\phi_\mu\phi^\mu+V_\phi(\phi)\right]\sqrt{-g}\nonumber\\
&-\frac{1}{G}\left[\frac{1}{2}g^{\mu\nu}\left(\frac{\nabla_\mu G\nabla_\nu G}{G^2}+\frac{\nabla_\mu\mu\nabla_\nu\mu}{\mu^2}-\nabla_\mu\nabla_\nu\omega\right)+\frac{V_G(G)}{G^2}+\frac{V_\mu(\mu)}{\mu^2}-V_\omega(\omega)\right]\sqrt{-g}.\nonumber
\end{align}
Here, $B_{\mu\nu}=\partial_\mu\phi_\nu-\partial_\nu\phi_\mu$, and $V_\phi(\phi)$, $V_G(G)$, $V_\omega(\omega)$, and $V_\mu(\mu)$ denote the self-interaction potentials associated with the vector field and the three scalar fields. The symbol $\nabla_\mu$ is used to denote covariant differentiation with respect to the metric $g^{\mu\nu}$, while the symbols $R$, $\Lambda$, and $g$ represent the Ricci-scalar, the cosmological constant, and the determinant of the metric tensor, respectively. We define the Ricci tensor as $R_{\mu\nu}=\partial_\alpha\Gamma^\alpha_{\mu\nu}-\partial_\nu\Gamma^\alpha_{\mu\alpha}+\Gamma^\alpha_{\mu\nu}\Gamma^\beta_{\alpha\beta}-\Gamma^\alpha_{\mu\beta}\Gamma^\beta_{\alpha\nu}$. Our units are such that the speed of light, $c=1$; we use the metric signature $(+,-,-,-)$.

The vector field is expected to produce a repulsive force. This is not possible unless matter carries a vector charge. Furthermore, the vector charge must have the right strength to cancel out excess gravity exactly such that the effective gravitational constant that remains is that of Newton. This means that the coupling term must also include a dependence on the scalar field $G$. This is important for another reason as well: a scalar charge is required in order to ensure that the theory survives precision solar system tests \cite{Moffat2010b}.

We specify this coupling in the case of a massive test particle by explicitly incorporating it into the test particle Lagrangian:
\begin{equation}
{\cal L}_\mathrm{TP}=-m+\alpha\omega q_5\phi_\mu u^\mu,
\end{equation}
where $\alpha$ is a function of $G$ and $q_5$ is the vector charge of a test particle with mass $m$ and four-velocity $u^\mu$.

This Lagrangian has been used in conjunction with the spherically symmetric, static vacuum solution of the MOG field equations to derive the phenomenology that we discuss in the next section.

\section{MOG phenomenology}
\label{sec:MOGpheno}

In MOG, the metric tensor is responsible for Einstein-like gravity, but $G$ is generally greater than Newton's constant, $G_N$.

The vector field is responsible for a repulsive force, canceling out part of the gravitational force; the effective gravitational constant at short range is $G_N$. The vector field is massive and has limited range; beyond its range, gravity is stronger than Newton predicts.

The strength of $G$ and the range $\mu^{-1}$ of the vector field are determined by the source mass.

In the weak field, low velocity limit, the acceleration due to a spherically symmetric source of mass $M$ is
\begin{equation}
\ddot{r}=-\frac{G_NM}{r^2}\left[1+\alpha-\alpha(1+\mu r)e^{-\mu r}\right],\label{eq:acc}
\end{equation}
where the overdot denotes differentiation with respect to time. The values of $\alpha$ and $\mu$ are determined by the source mass $M$ with formulas fitted using galaxy rotation and cosmology data:
\begin{align}
\alpha&=\frac{M}{(\sqrt{M}+E)^2}\left(\frac{G_\infty}{G_N}-1\right),&\mu=\frac{D}{\sqrt{M}},\\
D&\simeq 6250~M_\odot^{1/2}~\mathrm{kpc}^{-1},&E\simeq 25000~M_\odot^{1/2},&~~~~~~~~~~G_\infty=(1+\alpha)G_N\simeq 20G_N.\nonumber
\end{align}

This acceleration law is consistent with laboratory and solar system experiments, star clusters, galaxies, and galaxy clusters across (at least) 15 orders of magnitude. At short range, $\mu r\ll 1$, (\ref{eq:acc}) becomes Newton's gravitational acceleration law,
\begin{equation}
\ddot{r}\simeq-\frac{G_NM}{r^2},
\end{equation}
whereas at great distances, $\mu r\gg 1$, we get Newtonian gravity with an ``enhanced'' value of the gravitational constant,
\begin{equation}
\ddot{r}\simeq-(1+\alpha)\frac{G_NM}{r^2}.
\end{equation}

We also used this acceleration law to investigate the MOG predictions for the cosmic microwave background and the galaxy-galaxy matter power spectrum, which we discuss below.


\section{MOG and the CMB}
\label{sec:MOGCMB}

\begin{figure}
\includegraphics[width=0.75\linewidth,height=0.3\textheight]{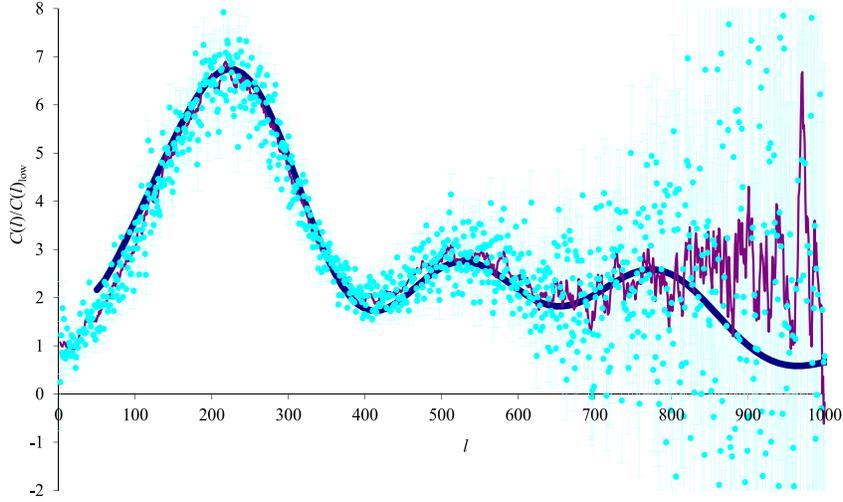}
\caption{The acoustic spectrum of the cosmic microwave background (5-year WMAP data points \cite{Komatsu2008} with error bars in light blue, with a moving window average also shown in purple) and the $\Lambda$CDM prediction (thick blue line).}
\label{fig:LCDMCMB}
\end{figure}

One of the key successes of the standard model of cosmology, $\Lambda$CDM, is its ability to predict the position and size of peaks in the acoustic power spectrum of the cosmic microwave background (Figure~\ref{fig:LCDMCMB}).

The question naturally arises: can MOG reproduce this result, especially in view of the fact that there exists no exotic dark matter in the MOG cosmology?

\begin{figure}
\includegraphics[width=0.65\linewidth,height=0.15\textheight]{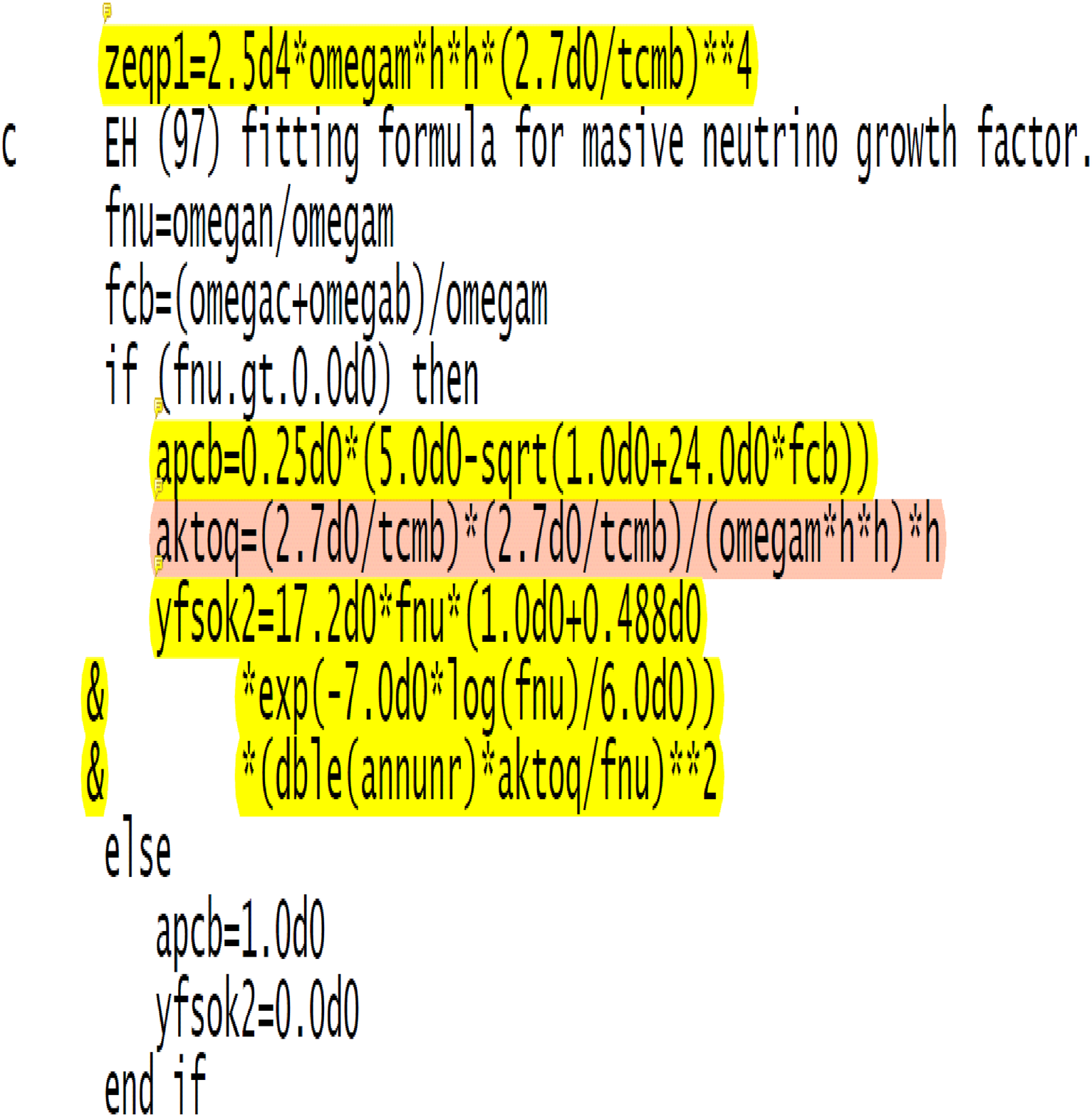}
\caption{Code fragment from the {\tt CMBFAST} \cite{Seljak1996} source distribution, demonstrating the difficulty of applying the program to a variable-$G$ cosmology, due to the fact that $\Omega\propto G\rho$ is used to represent constituent densities in both gravitational and non-gravitational contexts.}
\label{fig:CMBFAST}
\end{figure}

As we were attempting to address this question, colleagues often advised us to use the ``industry standard'' cosmological code {\tt CMBFAST} \cite{Seljak1996} or one of its derivatives such as {\tt CMBEASY} \cite{CMBEASY}. When we initiated a detailed study of these programs, however, we found that adapting them to a variable-$G$ cosmology is a highly nontrivial undertaking. A key reason is that CMBFAST uses variants of the cosmological quantity $\Omega$ (e.g., the baryon density $\Omega_b$) to represent matter in both gravitational (e.g., structure growth) and nongravitational (e.g., speed-of-sound calculations) contexts. In a variable-$G$ cosmology, $\Omega=8\pi G\rho/3H^2$ may change even as the corresponding density $\rho$ remains constant, due to changes in the value of $G$. Whereas gravitational relations involve the quantity $G\rho$, which $\Omega$ properly represents, nongravitational relations involve $\rho$.

This difficulty is by no means insurmountable, but it turns the adaptation of {\tt CMBFAST} into an arduous and error-prone task.

Other codes, such as {\tt CMBEASY}, often use a version of {\tt CMBFAST} as the underlying computational engine. Worse yet, the engine is often machine-translated from the original FORTRAN into another programming language, such as C or C++.

\begin{figure}
\includegraphics[width=0.75\linewidth,height=0.3\textheight]{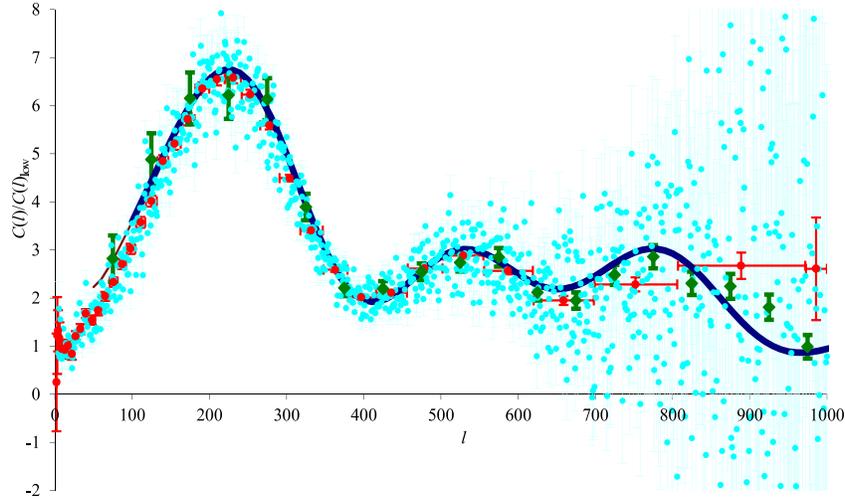}
\caption{The acoustic spectrum of the cosmic microwave background (WMAP data points with error bars in light blue) and the MOG prediction (thick blue line). Binned WMAP data (red) and Boomerang data (green) are also shown.}
\label{fig:MOGCMB}
\end{figure}

It was in part for this reason that we elected to take a closer look at a promising alternative: a semi-analytical approximation\footnote{It should be noted that the {\tt CMBFAST} code base also relies on semi-analytic formulations, e.g., those developed by Eisenstein and Hu \cite{EH1998}.} developed by Mukhanov \cite{Mukhanov2005} that is nonetheless more than just a collection of fitting formulae. Mukhanov's formulation does not hide the underlying physics, and it becomes a relatively straightforward substitution to replace, e.g., all occurrences of $\Omega$ with $(G_\mathrm{eff}/G_N)\Omega$ in gravitational contexts, where $G_\mathrm{eff}$ is the effective gravitational constant at the horizon. The result (Figure~\ref{fig:MOGCMB}) is encouraging but not altogether surprising: The enhanced gravitational constant plays the same role as collisionless dark matter in structure growth, but dissipation is due to the baryonic matter density, which is the same as in $\Lambda$CDM.

\section{MOG and the matter power spectrum}
\label{sec:MOGmatpow}

Another key prediction of $\Lambda$CDM cosmology, confirmed by observation, is the matter power spectrum: the statistic of density fluctuations in the large-scale distribution of galaxies.

This statistic can be understood using the Newtonian theory of small fluctuations in a self-gravitating medium. These fluctuations are governed by the equation
\begin{equation}
\ddot{\delta}_\vec{k}+\frac{\dot{a}}{a}\dot{\delta}_\vec{k}+\left(\frac{C_s^2k^2}{a^2}-4\pi G\rho\right)\delta_\vec{k}=0\label{eq:fluct}
\end{equation}
for each Fourier mode $\delta=\delta_\vec{k}(t)e^{i\vec{k}\cdot\vec{q}}$ (such that $\nabla^2\delta=-k^2\delta$; $C_s$ is the speed of sound in the medium). In the case of the standard model cosmology, $\nabla^2$ can be obtained from the Poisson equation of Newtonian gravity. In MOG, the acceleration law can be used to derive the corresponding inhomogeneous Helmholtz equation:
\begin{equation}
\nabla^2\Phi=4\pi G_N\rho(\vec{r})+
\alpha\mu^2G_N\int{\frac{e^{-\mu|\vec{r}-\vec{\tilde{r}}|}\rho(\vec{\tilde{r}})}{|\vec{r}-\vec{\tilde{r}}|}}~d^3\vec{\tilde{r}}.
\end{equation}

The Helmholtz equation leads a shifting of the wave number in the solution to (\ref{eq:fluct}):
\begin{equation}
k'^2=k^2+4\pi a^2\left(\frac{G_\mathrm{eff}-G_N}{G_N}\right)\lambda_J^{-2}.
\end{equation}

Changes to the sound horizon scale are unaffected by the varying strength of gravity. However, Silk damping introduces a $G^{3/4}$ dependence \cite{Padmanabhan1993}:
\begin{equation}
k'_\mathrm{Silk}=k_\mathrm{Silk}\left(\frac{G_\mathrm{eff}}{G_N}\right)^{3/4}.
\end{equation}

\begin{figure}
\includegraphics[width=0.75\linewidth,height=0.3\textheight]{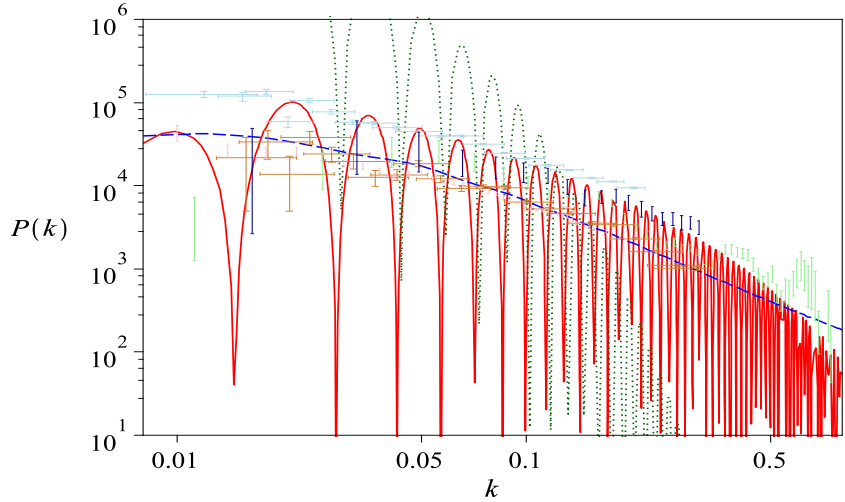}
\caption{The matter power spectrum. Three models are compared against five data sets: $\Lambda$CDM (dashed blue line, $\Omega_b=0.035$, $\Omega_c=0.245$, $\Omega_\Lambda=0.72$, $H=71$~km/s/Mpc), a baryon-only model (dotted green line, $\Omega_b=0.035$, $H=71$~km/s/Mpc), and MOG (solid red line, $\alpha=19$, $\mu=5h$~Mpc$^{-1}$, $\Omega_b=0.035$, $H=71$~km/s/Mpc.) Data points are colored light blue (SDSS 2006 \cite{SDSS2006}), gold (SDSS 2004 \cite{SDSS2004}), pink (2dF \cite{2dF2006}), light green (UKST \cite{UKST1999}), and dark blue (CfA \cite{CfA1994}).}
\label{fig:GGraw}
\end{figure}

These results can be substituted in the analytical approximations of Eisenstein and Hu \cite{EH1998}, leading to the the plot shown in Figure~\ref{fig:GGraw}.

Comparing the result visually against the data, we can tell that the solution appears to have the right slope; however, unlike the $\Lambda$CDM prediction, MOG predicts unit oscillations in the power spectrum. These unit oscillations are dampened in the case of $\Lambda$CDM by the presence of matter; in the case of MOG, no such matter is present and oscillations are not dampened.

\begin{figure}
\includegraphics[width=0.75\linewidth,height=0.3\textheight]{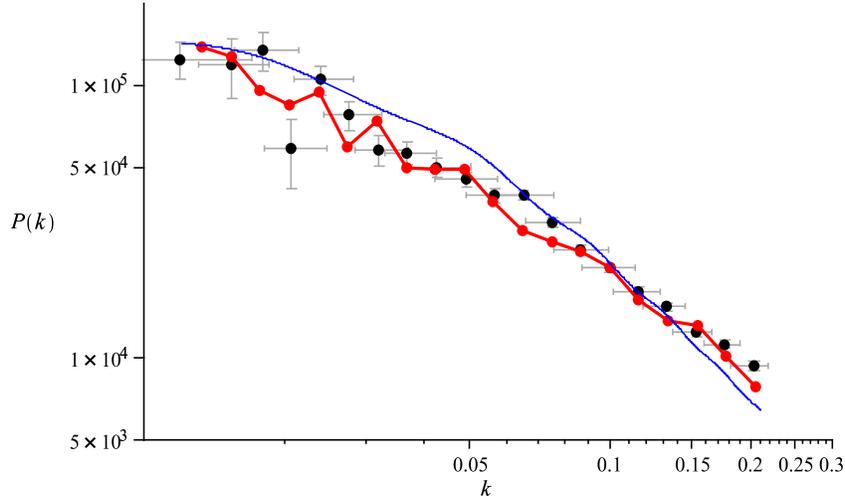}
\caption{The effect of window functions on the power spectrum is demonstrated by applying the SDSS luminous red galaxy survey window functions to the MOG prediction. Baryonic oscillations are greatly dampened in the resulting curve (solid red line). A normalized linear $\Lambda$CDM estimate is also shown (thin blue line) for comparison.}
\label{fig:GGwin}
\end{figure}

Are these unit oscillations present in the data? As it turns out, it is not yet possible to answer this question as the resolution of the data set is not sufficient. The data are effectively binned using a window function. When we apply that window function to the MOG prediction, the unit oscillations are smoothed out, and the result shows very good agreement indeed with the data points (Figure~\ref{fig:GGwin}).

In summary, two key features of the matter power spectrum are its slope and the presence or absence of baryonic oscillations. MOG reproduces the right slope; however, unlike $\Lambda$CDM, MOG has unit oscillations that are not dampened by the presence of collisionless dark matter. Future galaxy surveys will unambiguously show if unit oscillations are present in the data. Therefore, the matter power spectrum can be key to distinguish modified gravity without exotic dark matter from cold dark matter theories.

\section{MOG and continuous matter}
\label{sec:MOGfluid}

In the preceding sections, we discussed how MOG can be used to reproduce the observed characteristics of the CMB acoustic spectrum and the matter power spectrum. These results, however, were based on the MOG point particle solution (the point particle equation of motion in the presence of a spherically symmetric, static, vacuum gravitational field.)

The question naturally arises: Is it really appropriate to use the point particle solution for continuous distributions of matter? To answer this question, we have to consider an essential feature of MOG: namely that the strength of the (Newtonian) gravitational field and the range of the MOG repulsive vector force both vary as functions of the source mass. In other words, if we combine two masses together, the resulting field will not be a simple sum of the individual fields, not even in a crude approximation: the theory has an essential nonlinearity.

This means that we cannot blindly rely on the point particle solution to do MOG cosmology: it is necessary to solve the MOG field equations in the presence of continuous matter, such as a (perfect) fluid model. We have taken some tentative steps in this direction (see also \cite{Moffat2008a}).

How does MOG couple to continuous matter? The coupling must be consistent with two constraints:
\begin{itemize}
\item MOG must obey the weak equivalence principle (WEP) in order to not run afoul of many observations, e.g., E{\"o}tv{\"o}s-style experiments;
\item MOG must be compatible with precision solar system observations, expressed in the form of the parameters of the Parameterized Post-Newtonian (PPN) formalism, notably the Eddington-parameters $\beta$ and $\gamma$.
\end{itemize}

The two Eddington parameters $\beta$ and $\gamma$ determine deviations from the Newtonian potential in post-Newtonian models:
\begin{align}
g_{00}=&1-\frac{2M}{r}+2\beta\left(\frac{M}{r}\right)^2,\\
g_{0j}=&0,\\
g_{jk}=&-\left(1+2\frac{2\gamma M}{r}\right)\delta_{jk}
\end{align}
The Eddington-parameter $\beta$ is identically 1 for MOG. The Eddington-parameter $\gamma$ has the same value as in Jordan-Brans-Dicke theory \cite{Deng2009}, which can be ``cured'' by introducing a scalar charge that makes it conformally equivalent to the minimally coupled scalar theory \cite{Moffat2010b}.

The WEP is often interpreted as a requirement for a metric theory of gravity. This criteria is obviously not satisfied by MOG, as material particles are assumed to carry a vector charge and not move along geodesics determined by the metric. However, it is possible to consider a more relaxed interpretation of the WEP: a theory that satisfies the WEP observationally must be conformally equivalent to a metric theory of gravity. That is to say that there must exist a conformal transformation under which any non-minimal couplings between matter and non-metric gravity fields would vanish. The justification for this relaxed interpretation is that an observer, equipped with a classical instrument, would not be able to choose between conformally equivalent frames of reference: non-gravitational laws of physics, notably electromagnetism, are unaffected by a conformal transformation.

Conformal transformations add a vector degree of freedom (the special conformal transformation, a translation preceded and followed by an inversion, $x'^\mu=(x^\mu-b^\mu x^2)/(2-2b\cdot x+b^2x^2)$) and a scalar degree of freedom (dilation, $x'^\mu=\alpha x^\mu$); this agrees with the degrees of freedom to which the matter Lagrangian is expected to couple in the MOG theory. The metric tensor is conformally invariant up to a rescaling: $g'^{\mu\nu}=\alpha^{-2}(1-2b\cdot x+b^2x^2)^2g^{\mu\nu}$.

These considerations about the WEP and $\gamma$ can lead to a tentative general prescription for the coupling between the MOG fields and matter. We anticipate that the field equations for a perfect fluid will contain a vector charge in the form
\begin{equation}
\phi^\nu u_\nu J_\mu=\omega\frac{G-G_N}{G}T_\mu^\nu u_\nu,
\end{equation}
and a scalar charge in the form
\begin{equation}
GJ=-\frac{1}{2}T.
\end{equation}

Given an equation of state, we can now write down the MOG field equations in the case of the FLRW metric,
\begin{equation}
ds^2=dt^2-a^2(t)\left[(1-kr^2)^{-1}dr^2+r^2d\Omega^2\right].
\end{equation}
The equations are, after setting $\omega=\mathrm{const.}$,

\begin{align}
\left(\frac{\dot{a}}{a}\right)^2+\frac{k}{a^2}=\frac{8\pi G\rho}{3}-\frac{4\pi}{3}\left(\frac{\dot{G}^2}{G^2}+\frac{\dot{\mu}^2}{\mu^2}-\frac{1}{4\pi}G\omega\mu^2\phi_0^2\right)\nonumber\\
+\frac{2}{3}\omega GV_\phi+\frac{8\pi}{3}\left(\frac{V_G}{G^2}+\frac{V_\mu}{\mu^2}\right)+\frac{\Lambda}{3}+\frac{\dot{a}}{a}\frac{\dot{G}}{G},
\end{align}
\begin{align}
\frac{\ddot{a}}{a}=-\frac{4\pi G}{3}(\rho+3p)+\frac{8\pi}{3}\left(\frac{\dot{G}^2}{G^2}+\frac{\dot{\mu}^2}{\mu^2}-\frac{1}{4\pi}G\omega\mu^2\phi_0^2\right)\nonumber\\
+\frac{2}{3}\omega GV_\phi+\frac{8\pi}{3}\left(\frac{V_G}{G^2}+\frac{V_\mu}{\mu^2}\right)+\frac{\Lambda}{3}+H\frac{\dot{G}}{2G}+\frac{\ddot{G}}{2G}-\frac{\dot{G}^2}{G^2},
\end{align}
\begin{align}
\ddot{G}+3\frac{\dot{a}}{a}\dot{G}-\frac{3}{2}\frac{\dot{G}^2}{G}+\frac{G}{2}\left(\frac{\dot{\mu}^2}{\mu^2}\right)+\frac{3}{G}V_G-V_G'+G\left[\frac{V_\mu}{\mu^2}\right]+\frac{G}{8\pi}\Lambda-\frac{3G}{8\pi}\left(\frac{\ddot{a}}{a}+H^2\right)\nonumber\\
=-\frac{1}{2}G^2(\rho+3p),
\end{align}
\begin{align}
\ddot{\mu}+3\frac{\dot{a}}{a}\dot{\mu}-\frac{\dot{\mu}^2}{\mu}-\frac{\dot{G}}{G}\dot{\mu}+\frac{1}{4\pi}G\omega\mu^3\phi_0^2+\frac{2}{\mu}V_\mu-V'_\mu=0,\\
\omega\mu^2\phi_0-\omega\frac{\partial V_\phi}{\partial\phi_0}=4\pi J_0.
\end{align}

\begin{figure}
\includegraphics[width=0.75\linewidth,height=0.3\textheight]{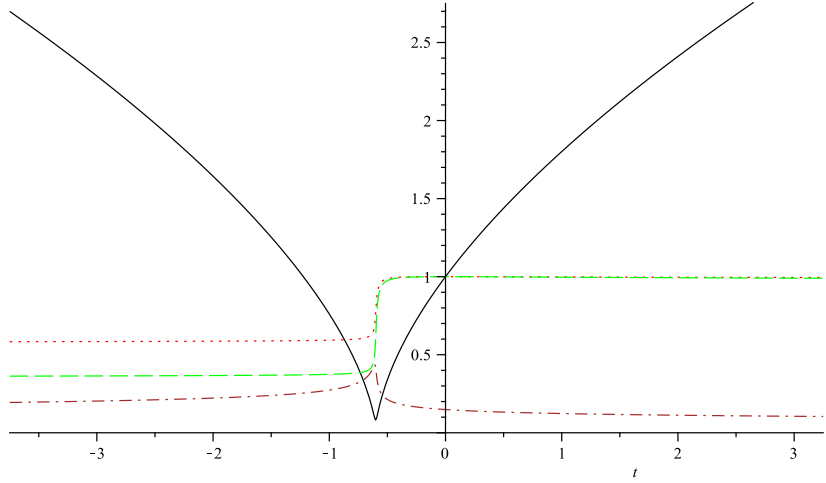}
\caption{The MOG ``classical bounce''. Black (solid) line is $a/a_0$ ; red (dotted) line is $G/G_0$ ; green (dashed) line is $\mu/\mu_0$ ; brown (dash-dot) line is $(a^3\rho)/(a_0^3\rho_0)$. Horizontal axis is in units of 13.7 billion years.}
\label{fig:bounce0}
\end{figure}

These FLRW field equations can be solved numerically, given suitable initial conditions and some assumptions. We generally ignore the self-interaction potentials:
\begin{equation}
V_\phi=V_G=V_\mu=0.
\end{equation}
We set the cosmological constant and the curvature constant to zero:
\begin{align}
\Lambda&=0,\\
k&=0.
\end{align}
We assume a simple equation of state, $p=w\rho$, and we are mainly interested in the late ``dust'' universe, $w=0$.

\begin{figure}
\includegraphics[width=0.75\linewidth,height=0.3\textheight]{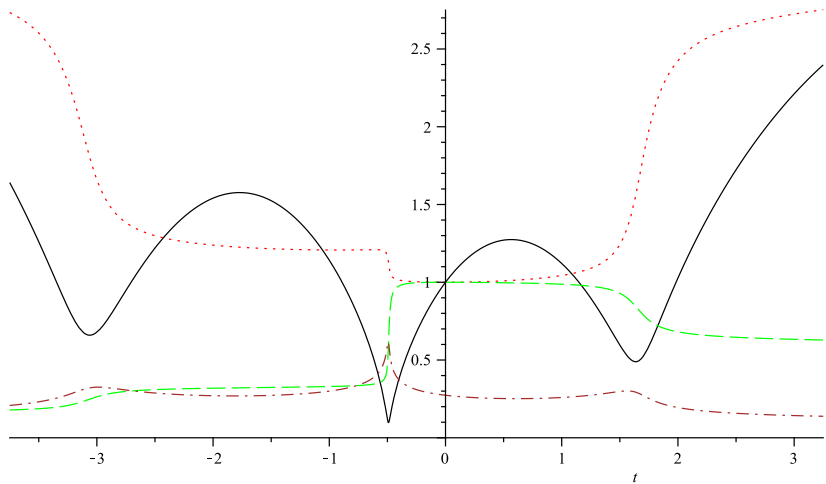}
\caption{The MOG ``cyclical bounce'' cosmology with negative $V_G=$~const. (For legend, see Figure~\ref{fig:bounce0}.)}
\label{fig:bounceneg}
\end{figure}

We use the present epoch to establish initial conditions: e.g., $\dot{a}/a|_{t=t_0}\simeq 2.3\times 10^{-18}~\mathrm{s}^{-1}$, and $\rho|_{t=t_0}\simeq 10^{-26}~\mathrm{kg}/\mathrm{m}^3.$ The solution yields a ``classical bounce'', albeit with an age problem (Figure~\ref{fig:bounce0}).

To address the age problem, we can consider using a non-zero value of $V_G$. We find that a negative value produces a cyclical universe, with repeated classical bounces (Figure~\ref{fig:bounceneg}). The conditions at the present epoch (notably, the expansion rate and approximate density of the universe) are repeated at later times during subsequent cycles. This allows for the possibility of a much older universe, one that has been through several cycles of shrinking and expansion since its densest state.

What about the deceleration parameter? This parameter, defined as $q=-\ddot{a}a/\dot{a}^2$, characterizes the rate at which the expansion rate slows. As it is well known, observations of the luminosity-distance relationship of distant Type Ia supernovae are inconsistent with a flat $q=0.5$ Einstein-de Sitter universe. The data may be consistent with an empty universe ($q=0$). It is also consistent with the $\Lambda$CDM universe that is dominated by dark energy ($\Omega_\Lambda\simeq 0.7$) at the present epoch.

\begin{figure}
\includegraphics[width=0.6\linewidth,height=0.12\textheight]{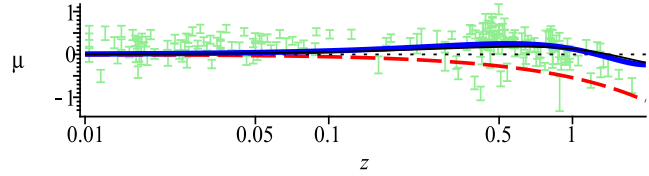}
\caption{Type Ia supernova luminosity-redshift data \citep{Riess2004} and the MOG/$\Lambda$CDM predictions. The horizontal axis corresponds to the $q=0$ empty universe. The MOG result is represented by a thick (blue) line. Dashed (red) line is a matter-dominated Einstein de-Sitter universe with $\Omega_M=1$, $q=0.5$. Thin (black) line is the $\Lambda$CDM prediction.}
\label{fig:TypeIa}
\end{figure}

MOG can also produce good agreement with the Type Ia supernova data if we assume the existence of a small positive value of $V_G$. This universe has a shallow bounce, and at the present epoch, its evolution is such that $\dot{a}/{a}$ is nearly constant. The actual age of the universe is, therefore, not fixed by the observed expansion rate alone; fitting to the supernova data yields a universe that is significantly older than the ``canonical'' value of 13.7 billion years. This model offers excellent agreement with the Type Ia supernova observations (Figure~\ref{fig:TypeIa}).

All of these models have shortcomings, but their existence shows that MOG is capable of producing physically plausible models of expansion, and that a classical bounce occurs naturally within the theory.

\section{Conclusions and outlook}
\label{sec:end}

The research of the cosmological consequences of MOG that is presented here is on-going. Important cosmological results, such as the matter power spectrum and the CMB acoustic spectrum, have been reproduced using the MOG equations of motion in the gravitational field of a point source. A general prescription that describes how MOG couples to matter (while remaining consistent with the WEP and precision observations) will allow us to solve the MOG field equations in the presence of matter \cite{Moffat2010a}, allowing us to re-derive the matter power spectrum and CMB acoustic spectrum results using a more solid foundation.

What can MOG tell us about the early universe? A bouncing cosmology may naturally avoid the horizon problem, which is a major motivation for inflationary cosmologies. It is unclear how MOG would address the flatness problem, other than postulating $k=0$ {\em a priori}.

What about Big Bang nucleosynthesis? At short range, the MOG acceleration law is consistent with Newton's. However, primordial isotope ratios are sensitive to the rate of cosmic expansion, which is governed by long-range gravity. We do not presently know if MOG can be consistent with observed isotope abundances.

Some of our results were obtained by assuming a constant value for $V_G$, the self-interaction potential for the $G$ scalar field. This assumption is {\em ad hoc}, not justified by theory. Perhaps improved solutions can lead us to eliminate the need to postulate a non-zero $V_G$, or alternatively, a theoretical justification for a (near) constant self-interaction potential can be found.

Notwithstanding these problems and open issues, MOG appears to be consistent with a range of astrophysical and cosmological phenomena. The theory yields a phenomenological acceleration law that works well across some 15 orders of magnitude in scale. Furthermore, there exists an unambiguous cosmological test through which MOG can be falsified: the absence of baryonic oscillations in the matter power spectrum cannot be explained by a MOG cosmology that lacks a cold dark matter component.




\newcommand{\mnras}{Mon. Not. R. Astron. Soc.}
\newcommand{\apj}{Astrophys. J.}
\newcommand{\prd}{Phys. Rev. D}
\newcommand{\jcap}{Journal of Cosmology and Astroparticle Physics}
\bibliography{refs}
\bibliographystyle{aipproc}   




%

\end{document}